\documentclass[aps, physrev, twocolumn, a4paper, superscriptaddress]{revtex4-1}

\usepackage{graphicx} 
\usepackage{dcolumn} 
\usepackage{amsmath,bm} 
\usepackage{color} 
\usepackage{gensymb} 

\begin{document}

\title{Magnetism and N\'{e}el skyrmion dynamics in GaV$_\mathbf{4}$S$_\mathbf{8-y}$Se$_\mathbf{y}$}

\author{T.~J.~Hicken}
\affiliation{Centre for Materials Physics, Durham University, Durham, DH1 3LE, United Kingdom}
\author{S.~J.~R.~Holt}
\affiliation{Department of Physics, University of Warwick, Coventry, CV4 7AL, United Kingdom}
\author{K.~J.~A.~Franke}
\affiliation{Centre for Materials Physics, Durham University, Durham, DH1 3LE, United Kingdom}
\affiliation{Department of Materials Science and Engineering, University of California, Berkeley, Berkeley, CA 94720, United States of America}
\affiliation{School of Physics and Astronomy, University of Leeds, Leeds, LS2 9JT, United Kingdom}
\author{Z.~Hawkhead}
\affiliation{Centre for Materials Physics, Durham University, Durham, DH1 3LE, United Kingdom}
\author{A.~{\v S}tefan{\v c}i{\v c}}
\affiliation{Department of Physics, University of Warwick, Coventry, CV4 7AL, United Kingdom}
\affiliation{Electrochemistry Laboratory, Paul Scherrer Institut, CH-5232 Villigen PSI, Switzerland}
\author{M.~N.~Wilson}
\affiliation{Centre for Materials Physics, Durham University, Durham, DH1 3LE, United Kingdom}
\author{M.~Gomil{\v s}ek}
\affiliation{Centre for Materials Physics, Durham University, Durham, DH1 3LE, United Kingdom}
\affiliation{Jo{\v z}ef Stefan Institute, Jamova c. 39, SI-1000 Ljubljana, Slovenia}
\author{B.~M.~Huddart}
\affiliation{Centre for Materials Physics, Durham University, Durham, DH1 3LE, United Kingdom}
\author{S.~J.~Clark}
\affiliation{Centre for Materials Physics, Durham University, Durham, DH1 3LE, United Kingdom}
\author{M.~R.~Lees}
\affiliation{Department of Physics, University of Warwick, Coventry, CV4 7AL, United Kingdom}
\author{F.~L.~Pratt}
\affiliation{ISIS Pulsed Neutron and Muon Facility, STFC Rutherford Appleton Laboratory, Harwell Oxford, Didcot, OX11 OQX, United Kingdom}
\author{S.~J.~Blundell}
\affiliation{Oxford University Department of Physics, Clarendon Laboratory, Parks Road, Oxford, OX1~3PU, United Kingdom}
\author{G.~Balakrishnan}
\affiliation{Department of Physics, University of Warwick, Coventry, CV4 7AL, United Kingdom}
\author{T.~Lancaster}
\affiliation{Centre for Materials Physics, Durham University, Durham, DH1 3LE, United Kingdom}

\date{\today}

\begin{abstract}
	We present an investigation of the influence of low-levels of chemical substitution on the magnetic ground state and N{\' e}el skyrmion lattice (SkL) state in GaV$_4$S$_{8-y}$Se$_y$, where $y =0, 0.1, 7.9$, and $8$.
	Muon-spin spectroscopy ($\mu$SR) measurements on $y=0$ and 0.1 materials reveal the magnetic ground state consists of microscopically coexisting incommensurate cycloidal and ferromagnetic environments, while chemical substitution leads to the growth of localized regions of increased spin density.
	$\mu$SR measurements of emergent low-frequency skyrmion dynamics show that the SkL exists under low-levels of substitution at both ends of the series.
	Skyrmionic excitations persist to temperatures below the equilibrium SkL in substituted samples, suggesting the presence of skyrmion precursors over a wide range of temperatures.
\end{abstract}
\maketitle

Chemical substitution is well-known for stabilizing exotic states of matter, from high-$T_\text{c}$ superconductivity in Mott insulators~\cite{lee2006doping}, to hidden magnetic order in heavy-fermion compounds~\cite{butch2010suppression,wilson2016antiferromagnetism,wilson2018muon}, and non-perturbative strongly-correlated Kondo states in itinerant systems~\cite{kondo1964resistance,hewson1997kondo,gomilvsek2019kondo}.
The influence of chemical substitution on topological defects, such as magnetic skyrmions~\cite{everschor2018perspective,lancaster2019skyrmions}, has been shown to be particularly pronounced, with substitution of just a few percent of the magnetic ions increasing the stability and lifetime of skyrmions~\cite{birch2019increased} and modifying their creation/annihilation energy barriers~\cite{wilson2019measuring}.
The study of the effects of low-levels of chemical substitution in bulk skyrmion materials has concentrated on materials hosting Bloch skyrmion lattice (SkL) states~\cite{muhlbauer2009skyrmion,munzer2010skyrmion,yu2011near,seki2012observation,kurumaji2019skyrmion}, as experimental realizations of N{\' e}el skyrmions in bulk materials are rare~\cite{kezsmarki2015neel,fujima2017thermodynamically,kurumaji2017neel,srivastava2019observation}.
We have previously studied the influence of high-levels of chemical substitution on the N{\' e}el SkL compounds GaV$_4$S$_8$ and GaV$_4$Se$_8$, by investigating $y=2$ and $4$ compositions in the GaV$_4$S$_{8-y}$Se$_y$ series~\cite{franke2018magnetic,stefancic2020establishing} and showed that it induces a spin-glass ground state and destroys the N{\' e}el SkL state.

Here we investigate the low-level limit of chemical substitution in the series through
muon spin spectroscopy ($\mu$SR)~\cite{blundell1999spin}, AC susceptibility~\cite{topping2018ac} and first principles calculations carried out using density functional theory (DFT)~\cite{clark2005first,perdew1996generalized,monkhurst1976special,muller2006magnetic,zhang2017magnetic}.
The influence of low-levels of substitutions on both the magnetic ground state and SkL in GaV$_4$S$_{8-y}$Se$_y$ is studied by comparing compositions very close to each end of the series, where $y=0.1$ and $y=7.9$, with N{\'e}el SkL-hosting $y=0$ (GaV$_4$S$_8$) and $y=8$ (GaV$_4$Se$_8$) compounds.
We find that the ground state of GaV$_4$S$_8$ is most sensitive to substitution, with significant changes of spin-density near substituents.
The dynamic signature of skyrmions persists in the substituted materials at both ends of the series, with an extended region of emergent low-frequency dynamics evident at low temperatures.

\begin{figure}
	\centering
	\includegraphics[width=\linewidth]{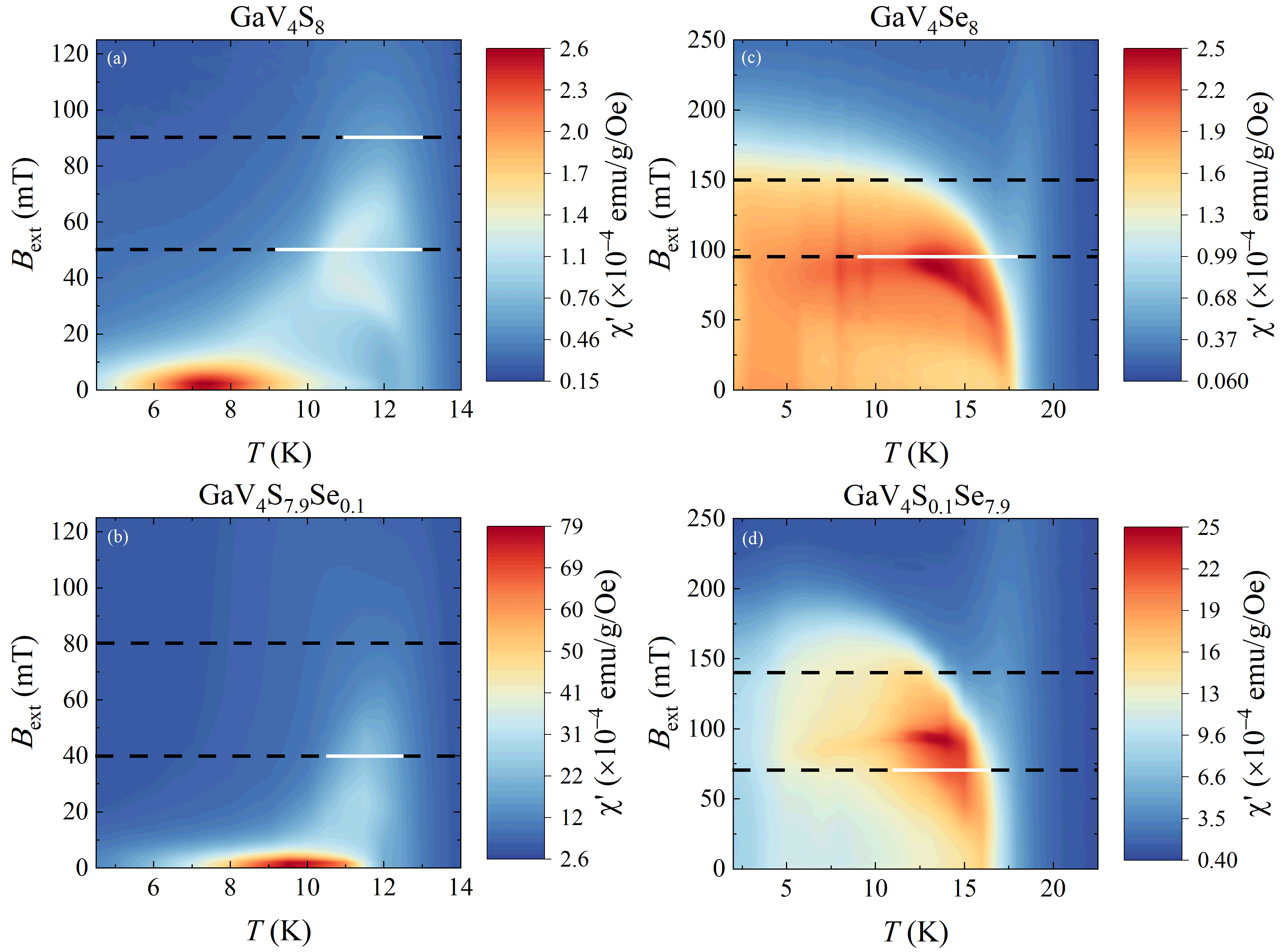}
	\caption{Real component of AC susceptibility in constant field $B_0$ for (a) $y=0$ (b) $y=0.1$, (c) $y=8$ and (d) $y=7.9$. Lines indicate fields where $\mu$SR measurements were performed, with white highlighting proposed SkL regions as based on $\mu$SR measurements (see main text).}
	\label{fig:ac}
\end{figure}

Polycrystalline samples of GaV$_4$S$_{8-y}$Se$_y$ were synthesized and characterized as described in Ref.~\cite{franke2018magnetic,stefancic2020establishing}.
Whilst polycrystalline samples are likely to exhibit a different magnetic phase diagram to those of single crystals (whose behavior varies depending on the alignment between applied field and crystallographic axes), we have previously shown that the SkL can be identified unambiguously~\cite{franke2018magnetic}.
AC magnetic susceptibility  (Fig.~\ref{fig:ac}) indicates relatively small changes in the position of the phase boundaries in the substituted materials when compared to their pristine counterparts, suggesting the SkL state is still formed.
In the substituted systems, the maximum susceptibility is approximately an order of magnitude greater than in the pristine systems, suggesting enhanced dynamics at low frequencies.

\begin{figure}
	\centering
	\includegraphics[width=\linewidth]{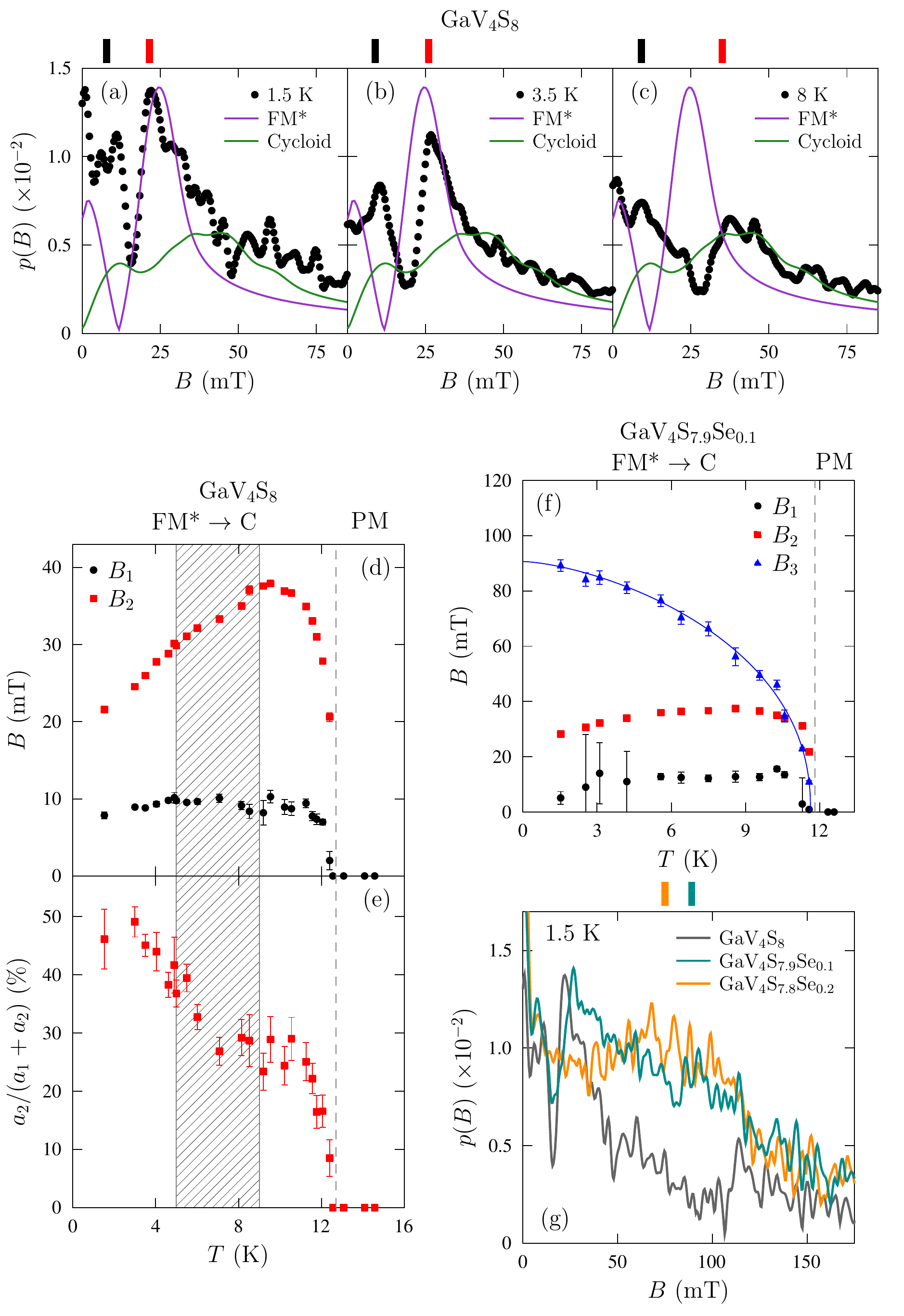}
	\caption{(a--c) Internal magnetic field distributions $p(B)$ for GaV$_4$S$_8$, obtained via the FTs of ZF $\mu$SR data at several temperatures, compared to simulations of the ferromagnetic-like (FM*) [Fig.~\ref{fig:strucdft}(a)] and cycloidal (C) states. Parameters from ZF $\mu$SR measurements of GaV$_4$S$_{8-y}$Se$_y$ for (d--e) $y=0$ and (f) $y=0.1$. The extracted internal fields seen in (d) are marked in (a--c). (g) $p(B)$ for $y=0$, $y=0.1$ and $y=0.2$, with the $B_3$ component marked for $y=0.1$ and $y=0.2$.}
	\label{fig:zffig}
\end{figure}

\begin{figure}
	\centering
	\includegraphics[width=\linewidth]{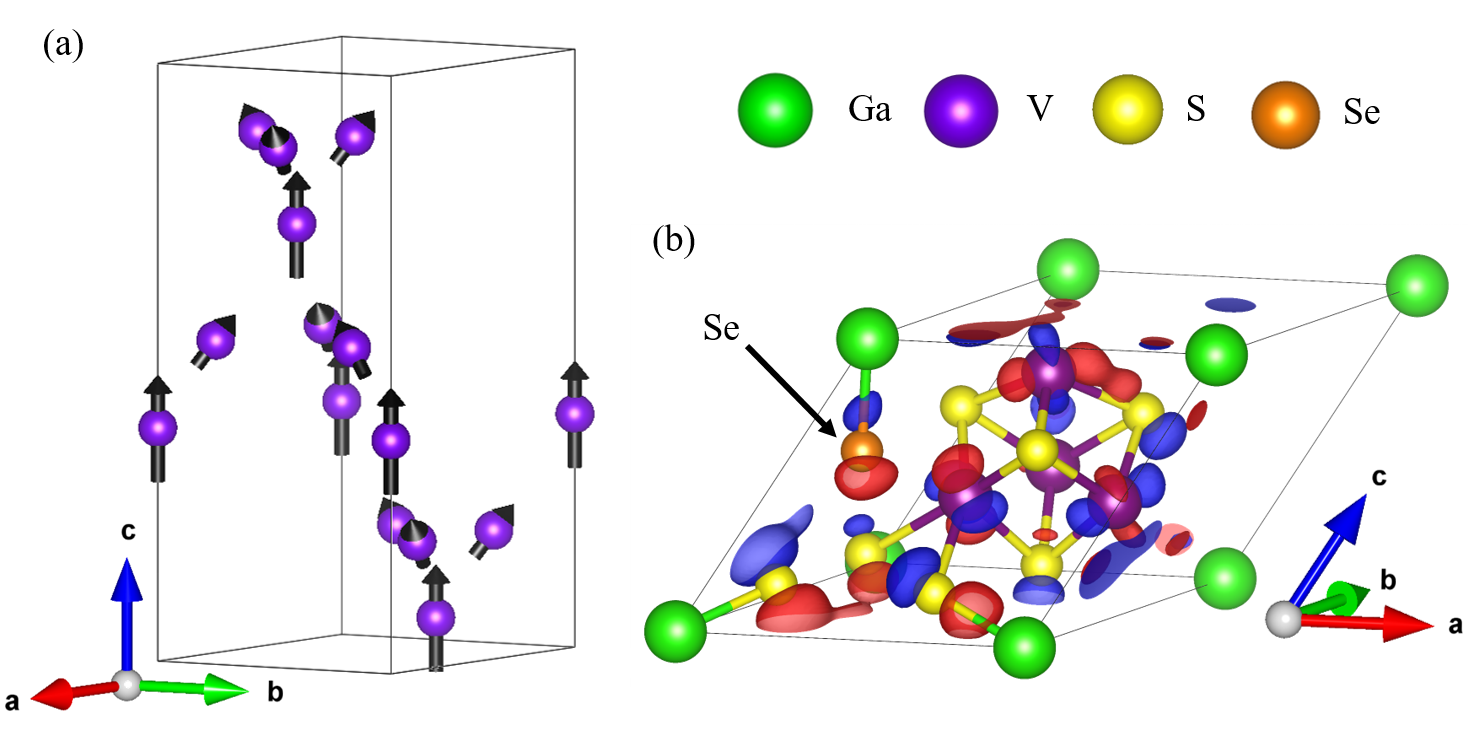}
	\caption{(a) FM* ground state for GaV$_4$S$_8$ (V atoms are shown.) (b) Difference in spin density between $y=0$ and $y=1$ for Se substitution on the S$_3$ site from DFT.}
	\label{fig:strucdft}
\end{figure}

To explore the magnetic ground state of the systems, zero-field (ZF) $\mu$SR measurements were performed.
In GaV$_4$S$_8$, above $T_{\mathrm{c}}~=~12.7(3)$~K we find the muon-spin polarization $P_z\left(t\right)$ is parameterized~\cite{pratt2000wimda} by $P_z\left(t\right) = a e^{-\lambda t} + a_\text{b} e^{-\lambda_\text{b} t}$, typical of a paramagnet.
The first term with amplitude $a$ reflects relaxation at rate $\lambda$ from muons that stop within the sample in the paramagnetic state, whilst the $a_\text{b}$ component captures the contributions from muons that stop outside the sample.
Fourier transforms (FTs) of $P_z\left(t\right)$ in the ordered phase ($T<T_{\mathrm{c}}$) are shown in Figs.~\ref{fig:zffig}(a--c).
Simulations of the magnetic field~\cite{bonfa2018introduction,si} at the muon-stopping sites~\cite{franke2018magnetic} for the ground-state magnetic structures proposed for GaV$_4$S$_8$~\cite{kezsmarki2015neel,ruff2015multiferroicity,stefancic2020establishing} show that the distribution most closely resembles the ferromagnetic-like (FM*) state [Fig.~\ref{fig:strucdft}(a)] at low-temperatures, and the incommensurate cycloidal (C) state at higher temperatures.
At all temperatures the spectra have features similar to those of both magnetic structures, however the data cannot be described by a simple sum of the two simulations as would be expected for spatially separated domains of FM* and C order.
Our data suggests a continuous evolution of the magnetic ground state from FM* to C (rather than an abrupt phase transition) where the spins slowly transform from one structure to the other.
In fact, this crossover has been suggested to occur via nucleation and growth of solitons~\cite{white2018direct,clements2020robust}, with the associated cycloidal anharmonicity likely to help explain some of the discrepancies between simulation and experiment.
The precise mechanism is likely to depend sensitively on the crystalline anisotropy in the system~\cite{izyumov1984modulated}.
Our data, therefore, suggests FM* domains are prevalent at low $T$, with the possibility of soliton-like cycloidal domain walls growing continuously with increasing $T$ until a C-majority phase is realized.

To compare the pristine and substituted systems at temperatures $T<T_{\mathrm{c}}$ the polarization is fitted to
\begin{equation}\label{eqn:asym_nosc}
P_z\left(t\right) = \sum_{i=1}^{n} a_i e^{-\Lambda_i t}\cos\left(\gamma_\mu B_i t + \phi_i\right) + a_\text{b} e^{-\lambda_\text{b} t} ,
\end{equation}
where each component with amplitude $a_i$, and relaxation rate $\Lambda_i$, reflects muons that stop in local field $B_i$ and precess with phase offset $\phi_i$.
For GaV$_4$S$_8$ we require only $n=2$, indicating two magnetically distinct components with local field magnitude $B_i$.
Extracted parameters [Figs.~\ref{fig:zffig}(d,e)] show that $B_1$ corresponds with the low-field peak seen in the simulations of both the FM* and C states (indicated in Figs.~\ref{fig:zffig}(a--c) with a black line), with $B_2$ corresponding to the high field peak (red line).
The unusual decrease in $B_{2}$ with decreasing temperature, along with a change in the fraction of muons subject to this magnetic field, reflects the continuous evolution of the magnetic state, providing further evidence for a smooth crossover between the FM* and C states, rather than a sharp phase transition \cite{franke2018magnetic}.
The crossover region $5\lesssim T \lesssim 9$~K [shaded in Fig.~\ref{fig:zffig}(e)] reflects the most rapid change of spin structure which leads to the enhanced AC susceptibility response seen in Fig.~\ref{fig:ac}(a).

For the GaV$_4$S$_{7.9}$Se$_{0.1}$ material, up to $T_{\text{c}}~=~11.6(2)$~K ZF $\mu$SR measurements are well parameterized by Eq.~\ref{eqn:asym_nosc} with $n=3$, indicating a third, magnetically-distinct muon environment not observed in GaV$_4$S$_8$ [Fig.~\ref{fig:zffig}(f)].
Below $T_\text{c}$, the amplitudes $a_i$ are found to be temperature independent, indicating that 13(5)\% of the muons stopping in the sample stop in sites with $B_1$, 32(3)\% in $B_2$, and 55(4)\% in $B_3$.
The ratio $n_1/n_2$ in GaV$_4$S$_{7.9}$Se$_{0.1}$ is consistent with the $T>8$~K region in GaV$_4$S$_8$ where C order dominates.
There are three mechanisms which can explain the appearance of the $B_3$ component in GaV$_4$S$_{7.9}$Se$_{0.1}$.
(i) A change in spin structure.
This can be ruled out as $B_1$ and $B_2$ are very similar in magnitude and $T$ evolution to GaV$_4$S$_8$, suggesting similar underlying behavior.
(ii) An increase in the magnetic moment $m$.
The field at the muon site $B_i \propto m$.
As $B_3/B_2 \simeq 3$ this would imply an increase of moment by the same factor, which can again be ruled out as there is no evidence for this in $B_1$ and $B_2$, or in DC magnetization measurements~\cite{si}.
(iii) A change in distance $r$ between the spin density and the muon.
As $B_i \propto 1/r^3$, even a modest change in spin density could lead to dramatic changes in $B_i$.
We therefore suggest that the most likely explanation of the appearance of the $B_3$ component is an increase in spin density near the muon sites such that these regions of high magnetic field condense around the substituent.
This is supported by the FT of $P_z\left(t\right)$ for GaV$_4$S$_{7.9}$Se$_{0.1}$ and GaV$_4$S$_{7.8}$Se$_{0.2}$ [Fig.~\ref{fig:zffig}(g)] which show that the signature FM* peak around 25~mT is further suppressed upon increased substitution, with spectral weight shifting to the broad, high-field peak not present in GaV$_4$S$_8$.
As the $B_3$ component becomes more pronounced with increased Se substitution this suggests the increase in spin density is not caused by muon implantation during the measurements.

The Se-rich end of the series shows more conventional magnetic behavior, with $\mu$SR measurements on the $y=8$ material suggesting $T_\text{c}~=~17.5(5)$~K with FT spectra consistent with simulations of a cycloidal spin structure~\cite{si}.
Measurements of GaV$_4$S$_{0.1}$Se$_{7.9}$ show similar average fields, but also feature an additional relaxing component and larger relaxation rates, indicating a broadening of the local magnetic-field distribution.

To further understand the effect of substitution we performed DFT calculations (see Ref.~\cite{si}) comparing the pristine materials to substitution of $y=1$ or $y=7$ (which are not measured here) by replacing the atoms on one S or Se site with the relevant substituent.
This allows us to simulate the effect of low-levels of substitution.
We compare the spin density in electron bands that dominate the contribution to the magnetism (i.e.\ those occupied in only one spin channel), with the difference between $y=0$ and $y=1$ for the S$_3$ site shown in Fig.~\ref{fig:strucdft}(b).
Regardless of the site chosen for substitution, more dramatic changes in spin density are seen at the S end of the series than at the Se end.
Whilst most of the change in spin density retains the expected d-orbital character, some site substitutions result in an increase in spin density within the V tetrahedral cluster. 
Since there are muon-stopping sites around the cluster, this provides an explanation for the observed large magnetic field $B_{3}$.
The change in spin density may also lead to altered exchange pathways, which upon increased substitution could lead to the glass-like magnetic ground state seen for $y=2$ and $y=4$~\cite{franke2018magnetic} where multiple different exchange pathways (depending on the local substitution level) cannot be simultaneously satisfied.

\begin{figure}
	\centering
	\includegraphics[width=\linewidth]{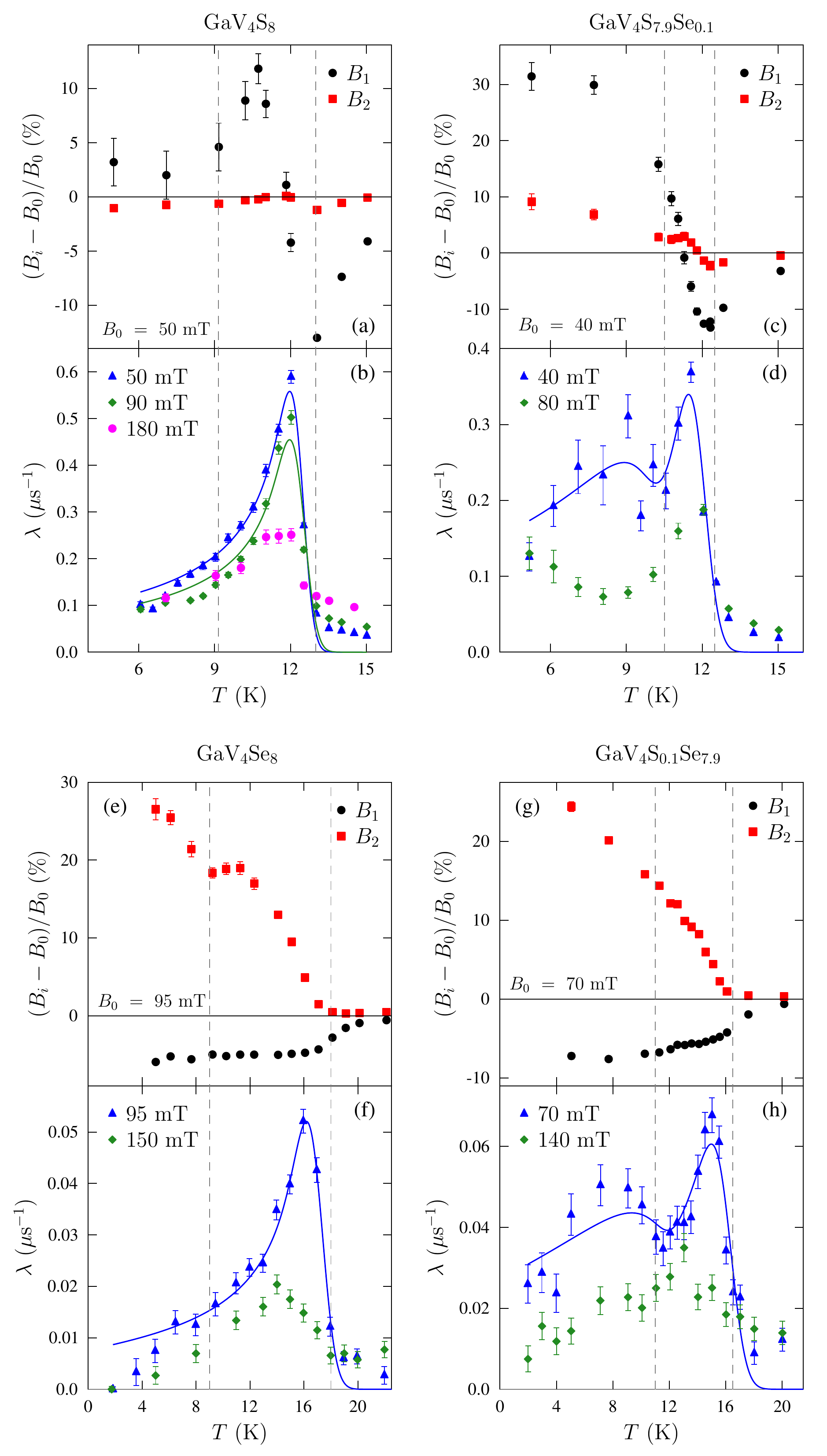}
	\caption{Parameters from TF $\mu$SR measurements on (a) $y=0$, (c) $y=0.1$, (e) $y=8$ and (g) $y=7.9$; and LF measurements for (b) $y=0$, (d) $y=0.1$, (f) $y=8$ and (h) $y=7.9$. Dashed lines are suggested boundaries for the SkL. Fits in (b), (d), (f) and (h) are detailed in the text.}
	\label{fig:skfig}
\end{figure}

We next explore the SkL state that appears in an applied field, through transverse-field (TF) and longitudinal-field (LF) $\mu$SR~\cite{si}.
Samples were cooled in zero applied magnetic field, and the measurements made in field (as indicated in Fig.~\ref{fig:ac}) on warming.
TF measurements are mainly sensitive to static disorder along with the component of dynamic fluctuations of the local field parallel to  the applied field, while LF measurements are sensitive to dynamics in those local fields perpendicular to the applied field.
The SkL orientation, determined predominantly by the crystalline anisotropy, will be randomized in a polycrystalline sample like ours, even under application of an external magnetic field, and hence the two techniques are expected to be sensitive to the same dynamic field correlations.
For the TF measurements the data is described by $P_x\left(t\right) = \sum_{i=1}^{2} a_i e^{-\Lambda_i t}\cos\left(\gamma_\mu B_i t + \phi_i\right) + a_\text{b}$, with results shown in Fig.~\ref{fig:skfig} and Ref.~\cite{si}.

In GaV$_4$S$_8$ a peak in $B_1$ is seen at $9~\lesssim~T~\lesssim~12~$K [Fig.~\ref{fig:skfig}(a)], coinciding with the presence of the SkL. 
This is consistent with $\mu$SR of the SkL in  materials such as Cu$_2$OSeO$_3$~\cite{lancaster2015transverse}, where an additional high-field shoulder is a signature of the SkL.
LF data for GaV$_4$S$_8$ are well parameterized by $P_z\left(t\right) \propto ae^{-\lambda t} + a_\text{b}$ over the entire temperature range, consistent with dynamic relaxation.
Measurements performed in an applied field of 50~mT and 90~mT cut through the SkL state in at least some crystal orientations~\cite{kezsmarki2015neel}, while measurements made at 180~mT do not.
These data [Fig.~\ref{fig:skfig}(b)] show that the effect of the dynamics in the SkL state is a significant enhancement in relaxation rate $\lambda$ below $T_{\mathrm{c}}$ leading to a large, broadened peak, centred at temperatures within the SkL state.
This is also consistent with LF measurements of the SkL in Cu$_2$OSeO$_3$~\cite{stefancic2018origin}.
We attribute the enhanced $\lambda$ to skyrmion excitation modes with frequency $\nu$ (such as the low-frequency rotational and breathing modes in the SkL plane~\cite{garst2017collective}) that soften (decreasing in frequency) as $T$ increases towards $T_\text{c}$~\cite{seki2017stabilization,tomasello2018micromagnetic}. 
As these modes cross through the frequency window where $\mu$SR is sensitive (around the Larmor resonance frequency $\omega_0 = \gamma_\mu B_\textrm{ext}$) the relaxation rate increases.
In the fast fluctuation regime $\lambda~=~2\Delta^2\nu/(\omega_0^2 + \nu^2)$ where $\Delta$ is the width of the local field at the muon site.
Applying power-law behavior typical for a 3D Heisenberg magnet $\nu~=~\nu_0\left(1-T/T_\text{c}\right)^{1.43}$ and $\Delta~=~\Delta_0\left(1-(T/T_\text{c})^{3/2}\right)^{0.365}$~\cite{blundell2003magnetism,pelka2018molecular,troyer1997critical,pospelov2019non,si}, produces good fits of $\lambda$, as seen in Fig.~\ref{fig:skfig}(b).
We find the zero-temperature skyrmion excitation mode frequency is approximately 10~GHz, consistent with the 3--17~GHz range observed in Bloch skyrmion materials~\cite{garst2017collective} and similar to the frequencies measured in single crystals of GaV$_4$S$_8$~\cite{ehlers2016skyrmion}.

The behavior of GaV$_4$Se$_8$ is similar to that of GaV$_4$S$_8$ with a peak observed in the TF field component $B_2$ [Fig.~\ref{fig:skfig}(e)].
LF $\mu$SR measurements on GaV$_4$Se$_8$ in an applied field of 95~mT, [Fig.~\ref{fig:skfig}(f)] also show a significantly enhanced relaxation rate in the SkL state (and peak below $T_{\mathrm{c}}$) when compared to the temperature scan with an applied field of 150~mT, where the SkL state is not stabilized.
This confirms the suggestion in Ref.~\cite{franke2018magnetic} that the SkL in polycrystalline samples is confined to a smaller region of the phase diagram than in single crystal samples.
A frequency of around 16~GHz for the zero-temperature excitation mode is found, again consistent with other skyrmion materials.

Finally, we turn to the influence of low-levels of substitution on the SkL state.
TF measurements for GaV$_4$S$_{7.9}$Se$_{0.1}$ were performed in an applied field of 40~mT, where an enhanced AC susceptibility response is consistent with a SkL state existing.
Unlike the GaV$_4$S$_8$ case, no unambiguous signature of the SkL state is observed in the internal magnetic field. 
In fact, the field  $B_1$ is significantly larger  for $y=0.1$ compared to $y=0$, consistent with the large internal field observed in our ZF $\mu$SR measurements.
This implies that any peak in internal field arising in the presence of the SkL state will be masked by these large fields.  
The variation in $B_2$ is similar between the samples, suggesting the underlying behavior is similar.
However, LF $\mu$SR measurements on the $y=0.1$ material [Fig.~\ref{fig:skfig}(d)] show a significantly enhanced peak in $\lambda$ at 40~mT compared to measurements at 80~mT, strongly suggestive of the characteristic dynamics of the SkL state.
This is accompanied by a region of enhanced $\lambda$ at lower temperature, observed only at fields where the SkL response is found.

At the Se-rich end of the series, fits of TF $\mu$SR measurements for $y=7.9$ in an applied field of 70~mT, where an enhanced AC susceptibility response consistent with a SkL is seen just below $T_\text{c}$, are shown in Fig.~\ref{fig:skfig}(g).
Although the overall trends in behavior are similar to those  for $y=8$, there is again no resolvable peak in internal field in the SkL region. 
However, LF $\mu$SR measurements [Fig.~\ref{fig:skfig}(h)] show that there is a clear enhanced response in $\lambda$ at 70~mT compared to 140~mT, consistent with the realization of the SkL state at 70~mT.
In addition, we again observe a separate enhancement in the low-temperature relaxation rate, similar to the behavior observed in $y=0.1$, with a broad peak in the relaxation rate centered around $T=8$~K.

The second, lower temperature peak in $\lambda$ that appears upon substitution only at fields at which the SkL is stabilized suggests that there are dynamics associated with the SkL extending down to lower temperatures.
We propose that these dynamics occur due to skyrmion precursors (as seen in Cu$_2$OSeO$_3$~\cite{bannenberg2019multiple}) that are stabilized by the subsituents at both ends of the series.
Note that if the dynamics were associated with the cycloidal phase one would expect a peak in $\lambda$ at all applied fields that stabilize the cycloidal phase.
The stabilization of the skyrmion precursors, either due to increased formation caused by to pinning, or to a longer lifetimes of metastable states, leads to dynamics detectable with $\mu$SR that are not dominant in the pristine materials.
These dynamics likely arise from a reduced frequency SkL mode due to the $\approx 300$ substituents per skyrmion present in the substituted systems, making the skyrmions less rigid, and hence lowering their characteristic frequencies.
Fitting this two-mode model for $\lambda$ [Figs.~\ref{fig:skfig}(d),(h)] reveals that the low-$T$ mode frequency is approximately 10--20\% of the high-$T$ frequency without a significant change in the width of the local field.

In conclusion, our results show that magnetic order is preserved for low-levels of substitution in GaV$_4$S$_{8-y}$Se$_{y}$, at both ends of the series, in contrast to higher levels where spin glass-like behavior is observed~\cite{franke2018magnetic,stefancic2020establishing}.
On the Se-rich end of the series, the presence of S simply leads to an increase in the width of the local magnetic field distribution, which is enough to prevent observation of the effect of the SkL in our TF measurements.
On the S-rich end of the series, where the ZF ordered state is formed from competing FM* and C order, substitution has a more dramatic effect, creating regions in the sample which have increased spin density, leading to the observation of high magnetic fields with $\mu$SR.
At both ends of the series LF $\mu$SR provides evidence of enhanced dynamics typical of those observed in N\'{e}el SkL states.
We therefore conclude that a dynamically fluctuating SkL is realized in these materials with low-levels of chemical substitution, with  skyrmion precursors at temperatures below the equilibrium skyrmion lattice.
We have shown further that the zero-temperature frequency of the N{\' e}el SkL excitation modes appear to be similar to those for a Bloch SkL, and suggest that skyrmion precursors may be ubiquitous over a wide range of temperatures in SkL materials.

Part of this work was carried our at the STFC ISIS Facility, UK and part at the Swiss Muon Source (S$\mu$S), Paul Scherrer Institut, Switzerland and we are grateful for the provision of beamtime. This project was funded by EPSRC (UK).
MG would like to acknowledge Slovenian Research Agency under project Z1-1852.
Research data from this paper will be made available via Durham Collections.

\bibliography{bib}

\end{document}